\documentclass[superscriptaddress,twocolumn,showpacs,preprintnumbers,prl]{revtex4-1}
\usepackage{times,xspace}
\usepackage{amsbsy,amssymb,amsmath,bm}
\usepackage{graphicx,color,epsfig}
\usepackage{fancyhdr}
\usepackage[usenames,dvipsnames,svgnames,table]{xcolor}
\usepackage{soul}

\def\bbbc{{\mathchoice {\setbox0=\hbox{$\displaystyle\rm C$}\hbox{\hbox
to0pt{\kern0.4\wd0\vrule height0.9\ht0\hss}\box0}}
{\setbox0=\hbox{$\textstyle\rm C$}\hbox{\hbox
to0pt{\kern0.4\wd0\vrule height0.9\ht0\hss}\box0}}
{\setbox0=\hbox{$\scriptstyle\rm C$}\hbox{\hbox
to0pt{\kern0.4\wd0\vrule height0.9\ht0\hss}\box0}}
{\setbox0=\hbox{$\scriptscriptstyle\rm C$}\hbox{\hbox
to0pt{\kern0.4\wd0\vrule height0.9\ht0\hss}\box0}}}}

\newcommand{\ignore}[1]{}
\newcommand{\mComment}[1]{}
\newcommand{\gComment}[1]{}
\newcommand{\jComment}[1]{}
\newcommand{\rComment}[1]{}
\newcommand{\lComment}[1]{}

\renewcommand{\gComment}[1]{\textcolor{magenta}{Gerardo: #1}}
\begin{document}

\title{Resistivity Minimum in Highly Frustrated Itinerant Magnets}

\author{Zhentao~Wang}
\affiliation{Department of Physics and Astronomy, Rice University, Houston, Texas 77005, USA}
\affiliation{Department of Physics and Astronomy, The University of Tennessee, Knoxville, Tennessee 37996, USA}
\author{Kipton~Barros}
\affiliation{Theoretical Division, T-4 and CNLS, Los Alamos National Laboratory, Los Alamos, New Mexico 87545, USA}
\author{Gia-Wei~Chern}
\affiliation{Department of Physics, University of Virginia, Charlottesville, Virginia 22904, USA}
\author{Dmitrii~L. Maslov}
\affiliation{Department of Physics, University of Florida, Gainesville, Florida 32611, USA}
\author{Cristian~D.~Batista}
\affiliation{Theory Division, T-4 and CNLS, Los Alamos National Laboratory, Los Alamos, New Mexico 87545, USA}
\affiliation{Department of Physics and Astronomy, The University of Tennessee, Knoxville, Tennessee 37996, USA}
\affiliation{Quantum Condensed Matter Division and Shull-Wollan Center, Oak Ridge National Laboratory, Oak Ridge, Tennessee 37831, USA}

 
\begin{abstract}
We study the transport properties of  frustrated itinerant magnets comprising localized {\it classical} moments, which interact
via exchange with the conduction electrons. Strong frustration stabilizes a liquidlike spin state, which extends down to
 temperatures well below the effective Ruderman-Kittel-Kasuya-Yosida interaction scale.
The crossover into this state is characterized by spin structure factor enhancement at wave vectors smaller than twice the Fermi wave vector magnitude. The corresponding enhancement of electron scattering generates a resistivity upturn at decreasing temperatures.
\end{abstract}
\maketitle

Certain magnetic metals  exhibit a resistivity minimum at low temperature. The Kondo effect explains this minimum 
via an effective exchange interaction $J$ between magnetic impurities and conduction electrons~\cite{Kondo64}.
Resistivity minima are also observed in  compounds comprising a periodic array of localized magnetic moments such as $4f$-electron compounds~\cite{Hewson97}. 
Because the Kondo effect is induced by spin-flip impurity scattering, 
it  is expected to be strongly suppressed in systems with large local magnetic moments or with strong easy-axis spin anisotropy.
Surprisingly, several compounds in this category,
such as Gd$_2$PdSi$_3$ and $R$CuAs$_2$ ($R$=Sm, Gd, Tb, and Dy)~\cite{Mallik1998_Gd2PdSi3, Sampathkumaran2003, Sengupta2004_RCuAs2}
and $R$InCu$_4$ ($R$=Gd, Dy, Ho, Er, and Tm)~\cite{Fritsch05,Fritsch06}, exhibit a pronounced resistivity minimum
despite heavy suppression of the Kondo effect.
These compounds are dominated by the Ruderman-Kittel-Kasuya-Yosida (RKKY) interaction, which competes against Kondo screening. 
It is natural to ask, therefore, if there exists a general mechanism by which a
RKKY interaction can induce a resistivity minimum~\cite{AA}.

In this Letter, we answer the question affirmatively:  frustrated itinerant magnets can exhibit a low-$T$ liquidlike spin state with enhanced resistivity
under quite general conditions.  For simplicity, we focus on a 2D Kondo lattice model (KLM)
with {\it classical}  local moments (no Kondo effect) and a small Fermi 
surface (FS). For a circular FS,  the bare magnetic  susceptibility $\chi^0_{\bm k}$ of the conduction electrons  
has a flat area of maxima for $k \leq 2 k_F$ (where $k \equiv |\bm k|$ and $k_F$ is the magnitude of Fermi wave vectors). 
The RKKY interaction thus seeks to enhance the structure factor (SF) in
the region $k \leq 2 k_F$.
We demonstrate that this effect
leads to an increase of
the electrical resistivity $\rho$ upon decreasing the temperature over the window $T_0 \lesssim T \lesssim | \theta_{\rm CW} |$,
where the magnetic correlation length increases from one lattice space $a$ (at $|\theta_{\rm CW}|$) to  $\xi \gg a$ (at $T_0$)~\footnote{Commonly, the magnetic correlation length $\xi$ is on the order of the lattice spacing at the Curie-Weiss temperature. However, it may develop at
higher temperatures in materials with competing ferro- and antiferromagnetic interactions. In this Letter, we 
define $|\theta_{\rm CW}|$ as the temperature at which $\xi$ becomes 
equal to the
lattice spacing. }.
Frustration ($| \theta_{\rm CW} |/T_0 \gg 1$) is required just to open this window; the rest is done by 
the nature of the RKKY interaction. The average enhancement of the spin SF
for wave vectors connecting points on the FS increases the elastic electron-spin scattering upon lowering $T$. 

The effect of the RKKY interaction on electron transport was considered in Refs.~\cite{Liu87} and \cite{Ruvalds88}. The sign of the effect
was found to be opposite (metallic) to that found in this Letter. This difference arises because we consider low filling, where the sign of $d\rho/dT$ can be shown to be insulating under quite general assumptions about the SF. 
In contrast, Refs.~\cite{Liu87} and \cite{Ruvalds88} considered a large FS,
 where the effect can have either sign depending on details of the electronic structure.

We first present an analytical derivation of the effect for the weak-coupling (WC) limit [$J  \eta(\varepsilon_F) \ll 1$, where $\eta(\varepsilon_F) $ is the density of states at the Fermi level]. The 
resistivity is evaluated in the Born approximation and the spin SF is obtained in two  ways: 
from a high-$T$ expansion~\cite{Oitmaa06} and by using the spherical approximation~\cite{Stanley68,Conlon10}.
Finally, we perform large-scale simulations of the full KLM. We use a variant of the
kernel polynomial method (KPM)~\cite{Weisse2006,Barros_PhysRevB.88.235101,Barros_PhysRevB.90.245119} to integrate Langevin dynamics (LD) and to evaluate the resistivity using the Kubo formula~\cite{Kubo1957}.
Our KPM-LD simulations on a triangular lattice (TL) with $256^2$ sites confirm the WC results and generalize them to the intermediate and strong-coupling regimes.

We consider the  KLM,
\begin{equation}
 {\cal H} \!= \!\! \sum_{\bm{k}, \sigma}  (\varepsilon_{\bm{k}}-\mu) c_{\bm{k}\sigma}^{\dagger}c_{\bm{k}\sigma}^{\;} \!
+ \! \frac{J}{\sqrt{N}} \!\! \sum_{{\bm q}, {\bm k}, \sigma, \sigma'} \!\!\!\!\!  c_{{\bm q}\sigma}^{\dagger}{\bm \sigma}_{\sigma\sigma'} c_{{\bm q}+{\bm k}\sigma'}^{\;} 
\! \cdot \! {\bm S}_{\bm k}.
\label{eq:Ham_KLM_k-sp}
\end{equation}
The operator $c_{{\bm k}\sigma}^\dagger (c_{{\bm k}\sigma}^{\;})$ creates (annihilates) 
an itinerant electron with momentum ${\bm k}$ and spin $\sigma$. 
$\varepsilon_{\bm k}= -\sum_{\boldsymbol \delta} t_{\boldsymbol \delta} e^{i {\bm k} \cdot {\boldsymbol \delta}}$ is the bare electronic dispersion relation with chemical potential $\mu$ and hopping amplitudes $t_{\boldsymbol \delta}$ between sites connected
by ${\boldsymbol \delta}$.  
The second term is the exchange interaction between the conduction electrons 
and the local magnetic  moments ${\bm S}_{\bm k}$ in  Fourier space.
We assume classical moments with magnitude $|\bm{S}_i| =1$ ($\bm \sigma$ is the vector of the Pauli matrices).

The conduction electrons can be integrated out in the WC limit
by expanding in the small parameter $J  \eta(\varepsilon_F)$. The resulting RKKY spin Hamiltonian is
\begin{equation}
H_\mathrm{RKKY} = - J^2 \sum_{\bm k} \chi^0_{\bm k} {\bm S}_{\bm k} \cdot {\bm S}_{\bar {\bm k}}
\label{RKKY}
\end{equation}
with ${\bar {\bm k}} \equiv -{\bm k}$ and ${\bm S}_{\bm k} = \sum_{l} e^{i {\bm k} \cdot {\bm r}_l} {\bm S}_l /\sqrt{N}$ ($N$ is the total number of lattice sites). 
The effective coupling constant in  momentum space is $-J^2 \chi^0_{\bm k}$ with $\chi^0_{\bm k} = T\sum_{\bm{q}, \omega_n} G^0_{\bm{q}, \omega_n} G^0_{\bm{q}+\bm{k}, \omega_n}$, 
where $\omega_n = (2n+1)\pi T$ are the Matsubara frequencies and
$G^0_{\bm{k}, \omega_n}= \left\{i \omega_n - \left[\varepsilon_{\bm k} - \mu\right]\right\}^{-1}$ 
is the bare Green's function. Then, the RKKY interaction favors magnetic orderings 
that maximize 
$\chi^0_{\bm k}$.

The electrons feel an effective potential produced by the spin configuration through the exchange interaction $J$.
If the system orders at low-enough temperature $(T \leq T_c)$,
the periodic array of spins only produces coherent electron scattering, which does not contribute 
to $\rho$~\footnote{Whether the resistivity drops or increases below $T_c$ depends on the competition of this effect and the reduction of the number of carriers due to the opening of a gap at the Fermi level. The resistivity behavior at $T_c$ is described in Ref.~\cite{Fisher1968}.}. 
However, the situation changes  above $T_c$ because the magnetic moments develop liquidlike correlations, which produce incoherent elastic electron-spin scattering. Within the Born approximation, the 
scattering cross section  is proportional to the spin SF,
\begin{equation}
{\cal S}({\bm k}) = \frac{1}{N} \sum_{jl} e^{i {\bm k} \cdot ({\bm r}_j - {\bm r}_l)} \langle {\bm S}_j \cdot {\bm S}_l \rangle = 
\langle {\bm S}_{\bm k} \cdot {\bm S}_{\bar {\bm k}} \rangle,
\end{equation}
where  $\langle \cdots \rangle$ denotes the thermodynamic average. 
${\cal S}({\bm k})$ satisfies the sum rule $\sum_{\bm k} {\cal S}({\bm k}) =N$ because $|\bm{S}_i| =1$.
Unlike the high-$T$ gas regime, characterized by a nearly ${\bm k}$-independent spin SF, 
short-range magnetic correlations appear in the liquid regime.  
The RKKY interaction is expected to enhance ${\cal S}({\bm k})$ for wave vectors connecting points on the FS because those are the processes that more effectively reduce the electronic energy. Given that the same processes contribute to the incoherent elastic scattering  in the paramagnetic state, $\rho$ should increase 
upon reducing  $T$ from the  high-$T$ gas regime to the $T_0 \lesssim T \lesssim |\theta_{\rm CW}| \sim J^2/t$ liquidlike regime.

To illustrate this point we will consider  the simple case of a circular FS,  relevant for most 2D lattices with a low electron
(hole) filling fraction~\footnote{Exceptions are fine-tuned systems with flat bands or lines of global maxima or minima of
$\varepsilon_{\bm k } $}. The  dispersion relation near the bottom (top) of the band can be approximated by
$\varepsilon_{\bm k} \simeq {\bm k}^2 / 2m$.  The resulting RKKY Hamiltonian is strongly frustrated: any spiral with wave vector ${\bm k}$ is a ground state as long as $k \leq 2 k_F$. The RKKY interaction favors these magnetic configurations because 
those are the only spirals that can scatter electrons between points ${\bm q}$ and ${\bm q}+{\bm k}$ on the FS.

Within the Born approximation, the inverse relaxation time for elastic scattering is
\begin{equation}
\frac{1}{ \tau_{k_F}} = \frac{4 \pi J^2}{N} \sum_{\bm k} \delta(\varepsilon_{F} - \varepsilon_{\bm k})
{\cal S}({\bm k} - {\bm k}_F)  (1 - \cos{\theta_{{\bm k}_F,{\bm k}}}).
\end{equation}
This expression is further simplified if ${\cal S}({\bm k} ) = {\cal S}(k)$, which is a good approximation
for low carrier filling fractions in the integration domain $k<2 k_F$:
\begin{equation}
\frac{1}{ \tau_{k_F}} = 4 \pi m  J^2 c  \int^{1}_0 dx \frac{x^2}{\sqrt{1-x^2}}\,
{\cal S}(2 k_F x),
\label{tau}
\end{equation}
where $c$ is a number that depends on the lattice, e.g., $c=\sqrt{3}/\pi^2$ for a TL.
The $T$ dependence of $\tau_{k_F}$ is then
determined by the variation of $ {\cal S}(k)$ for $k \leq 2 k_F$. 

\begin{figure}[t!]
	\includegraphics[width=\columnwidth]{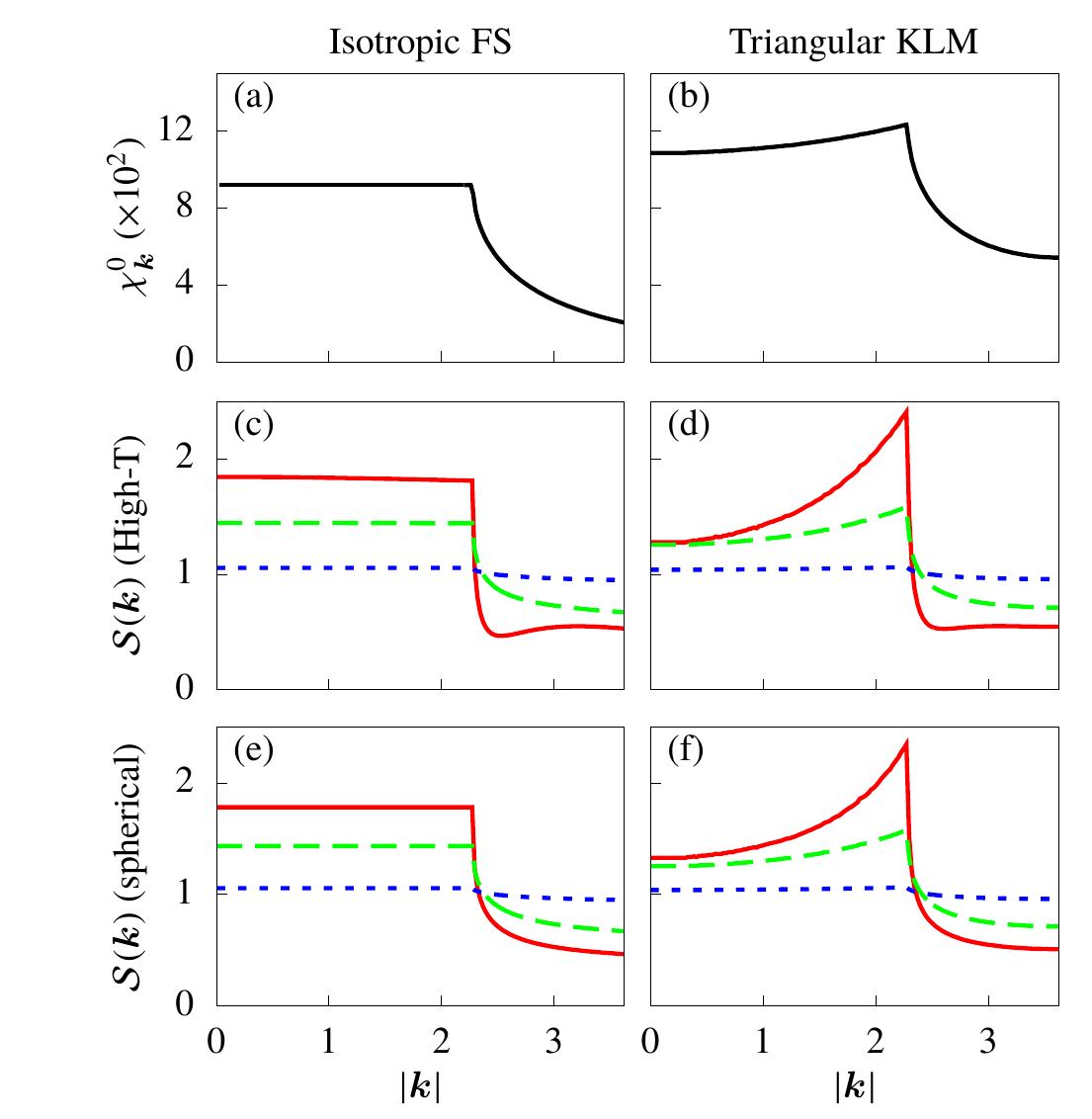}
	\caption{ \label{Fig1} 
Bare electronic susceptibility for (a) a 2D electron gas
	with isotropic dispersion $\varepsilon_{\bm k}=k^2/2m$, and (b) 
	a TL with NN hopping ($t=1$) and filling fraction $n=0.09$. 
	Panels (c)--(f) show the momentum dependence of  ${\cal S}({\bm k})$ at temperatures $T=\{ 0.03,0.06,0.45 \}J^2/t$ represented by solid, dashed, and dotted curves, respectively.
	Panels (c) and (d) are obtained from a high-$T$ expansion [see Eq.~\eqref{hte}], while panels (e) and (f) are obtained from the spherical approximation. Each panel is calculated using the bare magnetic susceptibility vertically above it.
	For panels (d) and (f) we assume  $\mathcal S({\bm k})\simeq \mathcal S(k)$, which is correct to within 1\% relative error.
    }
\end{figure}

We will use two 
independent approaches for computing the $T$ dependence of $ {\cal S}(k)$ in the  gas and liquidlike 
regimes. The first approach is a straightforward high-$T$ expansion~\cite{Oitmaa06}:
\begin{eqnarray}
{\cal S}( {\bm k} ) &=& 1 + K  {\tilde \chi}_{\bm k} + K^2
 \left [{\tilde \chi}^2_{\bm k} - \langle {\tilde \chi}^2 \rangle   \right] 
+ K^3 \big[  {\tilde \chi}^3_{\bm k} - \langle {\tilde \chi}^3 \rangle 
\nonumber \\
&& -2 {\tilde \chi}_{\bm k} \langle {\tilde \chi}^2 \rangle 
+ \frac{2}{5 N^2} \sum_{{\bm q} {\bm q}'}  
 {\tilde \chi}_{\bm q}   {\tilde \chi}_{\bm q'}   {\tilde \chi}_{{\bm k}-{\bm q}-{\bm q'}}  \big]
 \label{hte}
\end{eqnarray}
with $K = 2 J^2 \beta/3$, ${\tilde \chi}_{\bm k} = {\chi}^0_{\bm k} - \langle {\chi} \rangle$,
and $\langle {\tilde \chi}^n \rangle  = \sum_{\bm k} {\tilde \chi}^n_{\bm k}/N$.
Figure~\ref{Fig1}(a) shows the bare magnetic susceptibility for the isotropic FS under consideration.
Figure~\ref{Fig1}(b) shows the bare susceptibility for a TL with nearest-neighbor (NN) hopping
$t$ and an electron filling fraction $n=0.09$ (the mass is $m=1/3t$). As expected, the effect of the small $C_6$ lattice anisotropy 
(of order $k^6_F$) is to split the large global maxima degeneracy that would correspond to an isotropic ${\chi}^0_{\bm k}$. 
We will see that this splitting does not alter significantly the window of stability of the liquidlike regime.
Figures~\ref{Fig1}(c) and \ref{Fig1}(d) show the momentum dependence of the SF at different temperatures
obtained from Eq.~\eqref{hte} for the isotropic FS and the triangular KLM, respectively.

To understand the insulating sign of the temperature dependence of $1/\tau$, it suffices to analyze the second term in Eq.~(\ref{hte}), which gives the leading order
contribution to the momentum dependence of ${\cal S}( {\bm k} )$. Since $\chi_k>0$,
the prefactor of the $1/T$ term in $1/\tau$ is positive as long as the average of $\chi_k$  over the interval $(0,2k_F)$ in Eq.~(\ref{tau})  exceeds the contribution from $\langle\chi\rangle$, 
which is just a constant times $\langle\chi\rangle$. Suppose that $\chi^0_k$ does not vary dramatically in the interval $(0,2k_F)$, where it  can be estimated by some typical value $\bar \chi$, and falls off quickly for $k_F \ll k \ll b$, where $b\sim 1$ is the reciprocal lattice spacing. Then, the contribution of $\chi_k^0$ to the integral in Eq.~\eqref{tau} is on the order of $\bar\chi$. On the other hand, $\langle\chi\rangle$ is an average value over the entire Brillouin zone, normalized by its area. Therefore, $\langle\chi\rangle\sim\bar\chi (k_F/b)^2$, and the contribution from $\chi_k^0$ is reduced only by a small correction of order $(k_F/b)^2$~\cite{supp}. 

Compared with the high-$T$ expansion, the so-called spherical approximation~\cite{Stanley68,Conlon10} is less well controlled, but can be applied to a wider temperature range. The hard constraints $|{\bm S}_i| = 1$ are replaced with a global soft constraint $\sum_i |\bm S_i|^2 = N$, which renders the spin Hamiltonian quadratic and can be easily integrated to give
$
{\cal S}( {\bm k} ) = \frac{3T}{2 [ \Delta(T) - J^2 {\tilde \chi}_{\bm k}]}$,
where  $\Delta(T)$ is determined from the self-consistency equation~\cite{supp}:
$
\frac{1}{N} \sum_{\bm k} J^2/[\Delta(T) - J^2 {\tilde \chi}_{\bm k}] =  K.
\label{delta}
$
Figures~\ref{Fig1}(e) and \ref{Fig1}(f) show that the results  for the isotropic FS and the triangular KLM agree  with Figs.~\ref{Fig1}(c) and \ref{Fig1}(d) down to
$T \simeq 0.03 J^2/t$, at which point the high-$T$ expansion fails.

\begin{figure}[t!]
	\includegraphics[width=\columnwidth]{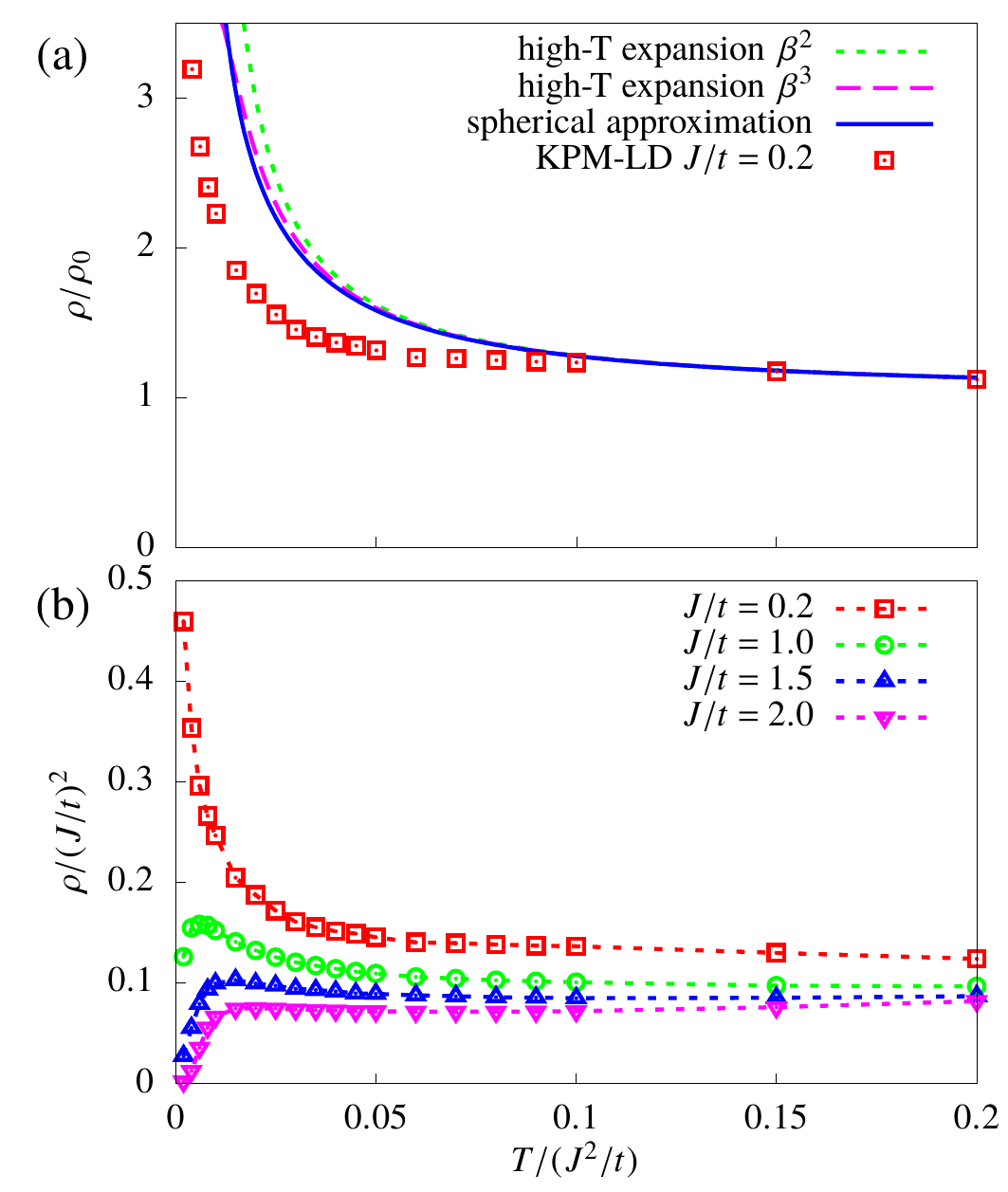}
	\caption{
	\label{Fig2} (a) Temperature dependence of the resistivity 
	for a triangular KLM with NN hopping ($t=1$) and  filling fraction $n=0.09$. 
	The lines correspond to calculations based on the the Born approximation [see Eq.~\eqref{rho}] and
	different analytical approaches for computing the temperature dependence of ${\cal S}({\bm k})$.
	The symbols correspond to the results of KPM-LD simulations rescaled by $\rho(T=J^2/t)$. (b) Resistivity curve (in units of $h/e^2$) obtained
	from KPM-LD simulations for different coupling strengths~\cite{comment_lowT}.
	}
\end{figure}

The electrical conductivity is given by
\begin{equation}
\sigma = -\frac{e^2}{2} \int \frac{\sqrt{3} d^2 k}{8 \pi^2} \tau_k v_k^2  \frac{d f(\varepsilon_k)}{d \varepsilon_k} 
\simeq \frac{3 \sqrt{3} e^2}{8 \pi} t k^2_F \tau_{k_F}.
\end{equation}
Replacing $\tau_{k_F}$ with its  expression given in Eq.~\eqref{tau}, we obtain
\begin{equation}
\rho(T) = \frac{4}{\pi}\rho_0 \int^{1}_0 dx \frac{x^2}{\sqrt{1-x^2}}
{\cal S}(2 k_F x),
\label{rho}
\end{equation}
where $\rho_0=8\pi J^2/(3tek_F)^2$.  Figure~\ref{Fig2}(a) compares the resistivity curves $\rho(T)$ obtained from 
the high-$T$ expansion and from the spherical approximation. As expected from the comparison
of the magnetic SF, the  resistivity curves practically
coincide down to $T \simeq 0.03 J^2/t$. Both curves 
confirm our main conjecture $d\rho/dT <0$  because the system develops 
stronger spin-spin correlations for wave vectors $k\leq 2 k_F$. 
This increase should be interrupted at $T=T_0$ where precursors of magnetic Bragg peaks develop from the broad peaks
of the liquid state and the Born approximation ceases to be valid.

The analytical approach that we have used for computing  $\rho(T)$  is only valid in 
the WC regime. Away from the WC regime, the RKKY theory is no longer valid as 
an effective low-energy theory for the KLM and the Born approximation is no longer justified.
Moreover, the two different approaches that we used for computing 
${\cal S}({\bm k})$ fail at low $T$. Our 
calculations then need to be complemented with numerical simulations valid for any coupling strength 
and down to arbitrarily low $T$. 

We perform KPM-LD simulations on a $256\times 256$ TL with small electron filling $n=0.09$ 
and
$J/t = (0.2,1.0,1.5, 2.0)$~\footnote{All lists of parameters in this paragraph correspond to the four values of $J/t$ in the parentheses.}. We integrate the dimensionless stochastic Landau-Lifshitz dynamics with a unit damping parameter using the Heun-projected scheme~\cite{Mentink10} for a total of $(2\times 10^3,4\times 10^3,6\times 10^3,1\times 10^4)$ time steps of duration $\Delta \tau =( 100,10,5,2 )$.
We estimate the effective spin forces 
using the gradient transformation described in Ref.~\onlinecite{Barros_PhysRevB.88.235101}. To decrease the stochastic error, we use the probing method of Ref.~\onlinecite{Tang12} with $R=128$ random vectors.
The Chebyshev polynomial expansion order is $M=500$.
To calculate the resistivity, we expand the Kubo-Bastin formula~\cite{Kubo1957,Bastin1971} using the KPM~\cite{Weisse2006,Garcia2015} with $M=(6000,1000,1000,500)$~\footnote{For $J/t=2$, the first three data points at low temperature are expanded up to $M=6000$.}. For each temperature, we average the longitudinal conductivity over ten snapshots separated by $(100,100,200,500)$ integration time steps.

Figure~\ref{Fig2}(b) shows the numerical $\rho(T)$ results for the different 
$J/t$ values. Frustration decreases with $J/t$ because higher order contributions (beyond RKKY level)
split the degeneracy for $k \leq 2 k_F$. For
the  strong-coupling limit $J \gg t$ the  
low-energy sector of ${\cal H}$ can be mapped into a double-exchange model, which favors 
ferromagnetic (FM) ordering at a critical temperature $T_c$ comparable to $|\theta_{\rm CW}|$. Given that the temperature window with liquidlike correlations diminishes as a function
of $J/t$, the relative low-temperature upturn of $\rho(T)$ should also decrease, as shown in
Fig.~\ref{Fig2}(b).

\begin{figure}[t!]
	\includegraphics[width=8.5cm]{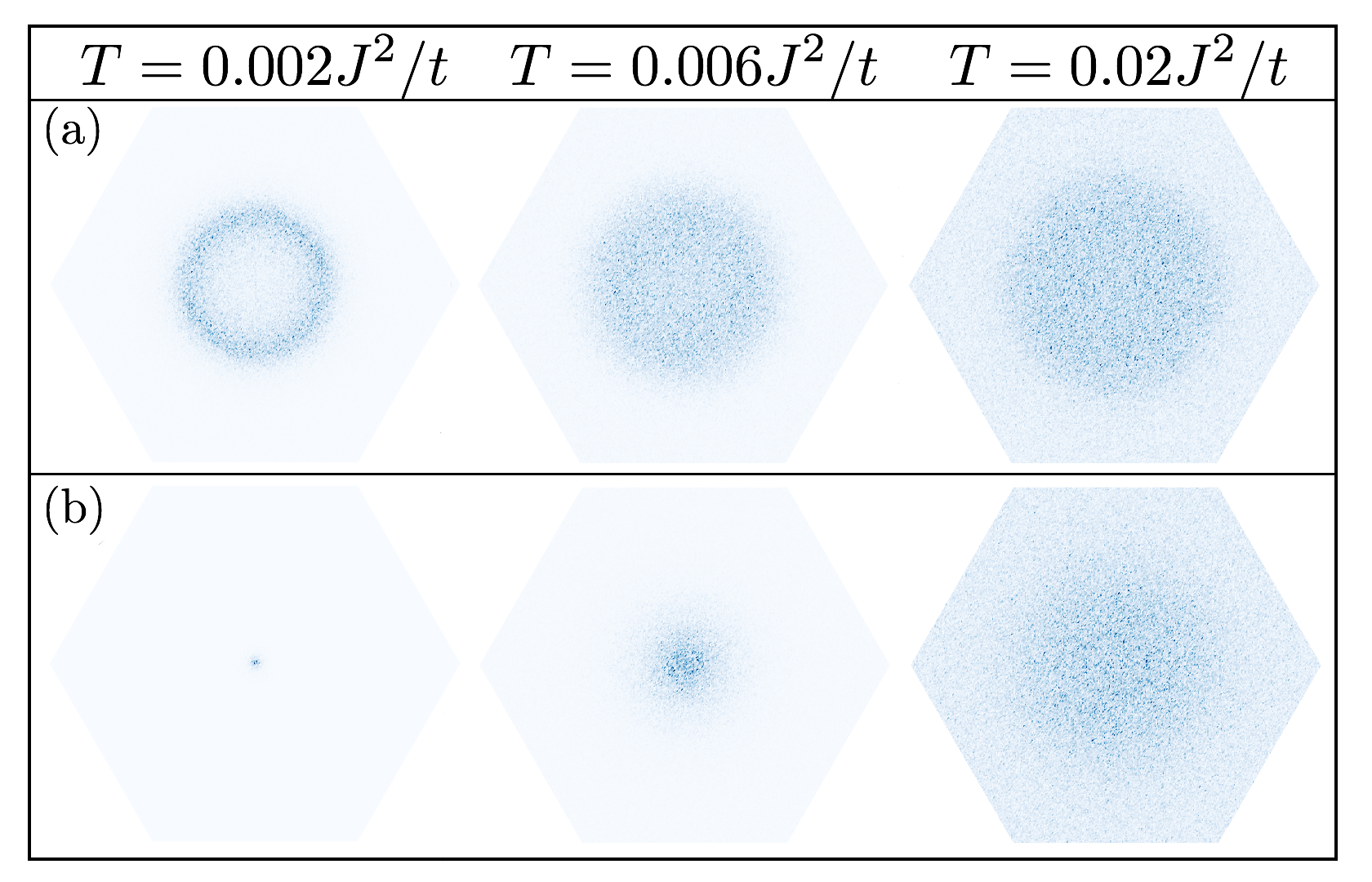}
	\caption{ \label{Fig3} 
Structure factor ${\cal S}({\bm k})$ for three temperatures at intermediate couplings
(a) $J/t=1$ and (b) $J/t=2$. At  $T \approx 0.02 J^2/t$,  ${\cal S}({\bm k})$
is nearly uniformly distributed in the disk $k\lesssim 2 k_F$. Around $T \approx 0.006 J^2/t$ the weight begins shifting toward a $k \lesssim 2 k_F$ radius ring ($J/t \approx 1$) or $k=0$ FM order ($J/t \gtrsim 2$). 
} 
\end{figure}

In the intermediate-coupling regime $J/t=1$, 1.5, and 2, the low-$T$ upturn of $\rho(T)$
reaches a maximum at temperature $T_0$  and drops rapidly for $T < T_0$.
 This crossover corresponds to the enhanced SF at wave vectors $k<2k_F$. Figures~\ref{Fig3}(a) and \ref{Fig3}(b) show the temperature dependence of ${\cal S}({\bm k})$ 
for $J/t = 1$ and $2$, respectively. The roughly uniform weight of ${\cal S}({\bm k})$ for $k<2k_F$ starts
redistributing below $T  \approx  0.006 J^2/t$. When $J/t \approx 1$ we observe the formation of a ring in Fourier space at $T \approx T_0$. 
This disordered phase is dynamically trapped at the lowest temperatures, $T \lesssim 0.002 J^2/t$.
As expected from the strong-coupling analysis, its radius $k_0 < 2 k_F$ decreases with $J/t$.
For larger couplings $J/t \gtrsim 2$ the FM phase clearly wins at low $T$.
We note that, for $T>T_0$, there is strong backward scattering produced by the $k \lesssim 2 k_F$ components of ${\cal S}({\bm k})$.
The resistivity drops below $T_0$ because the backscattering contribution ($k = 2 k_F$) 
is reduced by the formation of a ring at $k_0< 2 k_F$
[see the integrand of Eq.~\eqref{rho}].

Here, we have only considered the resistivity component arising from electron-spin scattering.  Electron-electron and electron-phonon scattering also contribute to $\rho$ in real materials.
These additional contributions increase with  $T$, whereas we have argued that the electron-spin scattering produces a negative $d\rho/dT$.  
The combination thus yields a resistivity minimum~\footnote{Except for cases when the electron-spin scattering is either too strong or too weak in comparison to the other scattering channels}.
Although we have assumed {\it classical} local spins ($S \rightarrow \infty$), our results can be extended to arbitrary $S$. The generalization of Eq.~\eqref{hte} is straightforward~\footnote{
The only difference is that the prefactors of $K^n$ become now functions of $S$.}.  
The main qualitative change is the Kondo effect expected for {\it quantum} spins and the antiferromagnetic exchange $J$.
This effect becomes apparent by applying the $T$-matrix formalism up to order $J^3$ to the KLM~\cite{supp}, 
which yields
\begin{equation}\label{Eq:kondo}
\rho(T) \approx \rho_{\text{RKKY}}(T)  \left[1-8J\eta(\epsilon_{F})\ln\left(\frac{k_{B}T}{D}\right)\right],
\end{equation}
where $\rho_{\text{RKKY}}(T)$ is given in Eq.~\eqref{rho}.
 $\rho_{\text{RKKY}}(T)$ becomes $T$ independent at $T \gg |\theta_{\rm CW}|$, so the only $T$ dependence arises from the Kondo effect.
According to Eq.~\eqref{Eq:kondo}, the Kondo logarithmic  behavior crosses over into a power law~\cite{supp}
\begin{equation}
\rho_\text{RKKY} (T) \sim \frac{a}{T-T^*}+b,
\end{equation}
upon entering the  range $T_0 \lesssim T \lesssim | \theta_{\rm CW} |$.
The qualitatively different $T$ dependence should allow us to distinguish between the two mechanisms for the resistivity upturn.
Moreover, the upturn produced by the RKKY mechanism should be accompanied by a corresponding upturn in the  correlation length $\xi$.
Indeed, moderately frustrated materials, such as Gd$_2$PdSi$_3$ and $R$CuAs$_2$ ($R$=Sm, Gd, Tb, and Dy)~\cite{Mallik1998_Gd2PdSi3, Sampathkumaran2003, Sengupta2004_RCuAs2}, exhibit a nonlogarithmic  resistivity upturn right above the N{\'e}el temperature. According to Refs.~\cite{Udagawa12,Chern13}, the resistivity minimum of the pyrochlore oxides Pr$_2$Ir$_2$O$_7$ and Nd$_2$Ir$_2$O$_7$ is also caused by spin-spin correlations described by the spin ice model.

Furthermore, the Kondo effect is absent in transition metal oxides, where $J$ is FM (Hund's coupling). Our results indicate that the resistivity upturn persists in the intermediate coupling regime, relevant to these materials.
Indeed, a resistivity upturn has been observed in (Ga$_{1-x}$Mn$_x$)As~\cite{Matsukura1998,Jungwirth2006} and manganites~\cite{Salamon2001} above the FM transition temperature $T_c$.

Our key conclusion is that the  RKKY interaction enhances the elastic electron-spin scattering by increasing  the  magnetic SF for  wave vectors connecting  points on
the FS. Assuming  that this enhancement eventually leads to Bragg peaks (for $T<T_c$), which do not
produce incoherent scattering, frustration is necessary to open a wide enough temperature window (liquidlike regime) over which the resistivity upturn becomes noticeable.
Although we have focused on 2D systems with a small FS,
the conclusion applies generally to  frustrated itinerant magnets, provided that  $\chi^0_{\bm k}$
is larger on average for wave vectors $\bm k$ connecting points on the FS.

\begin{acknowledgments}
We thank  A. Chubukov, S. Maiti, F. Ronning, E. V. Sampathkumaran, and J. D. Thompson for useful discussions. 
Z.W. acknowledges support from the CNLS summer student program and Welch Foundation Grant No.~\mbox{C-1818}.
Computer resources for numerical calculations were supported by the Institutional Computing Program at LANL. This work was carried out under the auspices of the NNSA of the U.S. DOE at LANL under Contract No. DE-AC52-06NA25396, and was supported by the U.S. Department of Energy, Office of Basic Energy Sciences, Division of Materials Sciences and Engineering. D.L.M. acknowledges support from the
National Science Foundation via Grant No. NSF DMR-1308972 and a Stanislaw Ulam
Scholarship at the CNLS, LANL.
\end{acknowledgments}

\setcounter{figure}{0}
\renewcommand{\thefigure}{S\arabic{figure}}
\setcounter{equation}{0}
\renewcommand{\theequation}{S\arabic{equation}}

\newpage
\begin{center}
  {\bf ---Supplemental Material---}
\end{center}

\section{Perturbation Theory for KLM}
For a Kondo lattice model (KLM), the temperature dependence of the resistivity  has contributions from {\it single} spin-flip scattering  (Kondo effect), and  from multi-spin scattering processes, which are controlled by the RKKY mechanism discussed in this paper. Below, we show how both effects  contribute to the transport to lowest non-trivial order in perturbation theory.

We consider the KLM in the main text $\mathcal{H}=\mathcal{H}_0 + \mathcal{H}_1$, and allow the local moments to be quantum mechanical spins:
\begin{subequations}\label{eq_supp:Ham_KLM}
\begin{align}
\mathcal{H}_0 &=\sum_{\bm{k}, \sigma}  (\varepsilon_{\bm{k}}-\mu) c_{\bm{k}\sigma}^{\dagger}c_{\bm{k}\sigma}^{\;}, \\
\mathcal{H}_1 &=\frac{J}{\sqrt{N}}  \sum_{{\bm q}, {\bm k}, \sigma, \sigma'}  c_{{\bm q}\sigma}^{\dagger}{\bm \sigma}_{\sigma\sigma'} c_{{\bm q}+{\bm k}\sigma'}^{\;}  \cdot  {\bm S}_{\bm k}.
\end{align}
\end{subequations}

The inverse relaxation time is obtained from the T-matrix formalism:
\begin{equation}
\frac{1}{\tau_{\bm{q}}} = 2\pi \sum_{\bm{k}} \delta (\epsilon_{\bm{q}}-\epsilon_{\bm{k}}) \sum_{\sigma\sigma^{\prime}} \left|T_{\bm{q}\sigma,\bm{k}\sigma^{\prime}}\right|^{2} \left(1-\cos\theta_{\bm{q},\bm{k}}\right),
\end{equation}
where the T-matrix is 
\begin{equation}
T = \mathcal{H}_1 + \mathcal{H}_1 G_0  \mathcal{H}_1,
\end{equation}
up to second order in $J$
and $G_0$ is the non-interacting electronic Green's function.
More explicitly:
\begin{subequations}
\begin{align}
T_{\bm{q}\uparrow,\bm{k}\uparrow} &=
\frac{J}{\sqrt{N}}S_{\bm{k}-\bm{q}}^{z} +\frac{J^{2}}{N}\sum_{\bm{p}} \Bigg[
-\frac{2}{\sqrt{N}}S_{\bm{k}-\bm{q}}^{z}\frac{1-f(\epsilon_{\bm{p}})}{\epsilon_{\bm{k}}-\epsilon_{\bm{p}}}  \nonumber \\
&\quad + \frac{S_{\bm{p}-\bm{q}}^{z}S_{\bm{k}-\bm{p}}^{z}+S_{\bm{k}-\bm{p}}^{+}S_{\bm{p}-\bm{q}}^{-}}{\epsilon_{\bm{k}}-\epsilon_{\bm{p}}} 
\Bigg], \\
T_{\bm{q}\downarrow,\bm{k}\downarrow} &=
-\frac{J}{\sqrt{N}}S_{\bm{k}-\bm{q}}^{z} +\frac{J^{2}}{N}\sum_{\bm{p}} \Bigg[ -\frac{2}{\sqrt{N}}S_{\bm{k}-\bm{q}}^{z}\frac{f(\epsilon_{\bm{p}})}{\epsilon_{\bm{k}}-\epsilon_{\bm{p}}} \nonumber \\
&\quad +\frac{S_{\bm{p}-\bm{q}}^{z}S_{\bm{k}-\bm{p}}^{z}+S_{\bm{p}-\bm{q}}^{+}S_{\bm{k}-\bm{p}}^{-}}{\epsilon_{\bm{k}}-\epsilon_{\bm{p}}}\Bigg],\\
T_{\bm{q}\downarrow,\bm{k}\uparrow} &=
\frac{J}{\sqrt{N}}S_{\bm{k}-\bm{q}}^{+} +\frac{J^{2}}{N}\sum_{\bm{p}} \Bigg[ \frac{1}{\sqrt{N}}S_{\bm{k}-\bm{q}}^{+}\frac{2f(\epsilon_{\bm{p}})-1}{\epsilon_{\bm{k}}-\epsilon_{\bm{p}}} \nonumber \\
&\quad +\frac{S_{\bm{p}-\bm{q}}^{+}S_{\bm{k}-\bm{p}}^{z}-S_{\bm{k}-\bm{p}}^{+}S_{\bm{p}-\bm{q}}^{z}}{\epsilon_{\bm{k}}-\epsilon_{\bm{p}}}\Bigg], \\
T_{\bm{q}\uparrow,\bm{k}\downarrow} &=
\frac{J}{\sqrt{N}}S_{\bm{k}-\bm{q}}^{-} +\frac{J^{2}}{N} \sum_{\bm{p}} \Bigg[ \frac{1}{\sqrt{N}}S_{\bm{k}-\bm{q}}^{-}\frac{2f(\epsilon_{\bm{p}})-1}{\epsilon_{\bm{k}}-\epsilon_{\bm{p}}} \nonumber \\
&\quad +\frac{S_{\bm{k}-\bm{p}}^{-}S_{\bm{p}-\bm{q}}^{z}-S_{\bm{p}-\bm{q}}^{-}S_{\bm{k}-\bm{p}}^{z}}{\epsilon_{\bm{k}}-\epsilon_{\bm{p}}}\Bigg],
\end{align}
\end{subequations}
where $f(\epsilon)$ is the Fermi distribution function.

For elastic scattering in the paramagnetic phase, it follows that
\begin{equation}
\sum_{\sigma\sigma^{\prime}} \! \left|T_{\bm{q}\sigma,\bm{k}\sigma^{\prime}}\right|^{2} \!\! = \!\! \frac{2J^{2}}{N}\mathcal{S}(\bm{k}-\bm{q}) \!\! \left[1 \! + \! \frac{2J}{N} \! \sum_{\bm{p}} \! \frac{2f(\epsilon_{\bm{p}})-1}{\epsilon_{\bm{k}}-\epsilon_{\bm{p}}}\right],
\end{equation}
where $\mathcal{S}(\bm{k})  = \langle \bm{S}_{\bm{k}} \cdot \bm{S}_{\bar{\bm{k}}}  \rangle$ is the spin structure factor.

The resistivity is then given by
\begin{equation}
\rho(T) \approx \rho_{\text{RKKY}}(T) \cdot \left[1- 8J\eta(\epsilon_{F})\ln\left(\frac{k_{B}T}{D}\right)\right],
\end{equation}
where $\eta(\epsilon_{F})$ is the density of states at Fermi level, and $D$ is the bandwidth. The logarithmic contribution of order $J^3$ corresponds to the celebrated Kondo effect.
The temperature dependence of the prefactor, $\rho_{\text{RKKY}}(T)$, arises from multi-spin scattering processes. To leading order (Born approximation), this temperature dependence is essentially the temperature dependence of the structure factor for wave-vectors connecting points on the Fermi surface. Back scattering processes ($k=2k_F$) are the ones that have the largest weight in Eq.~(8) of the main text.



\section{Sign of the temperature dependence of $1/\tau$}

According to Eqs.~(5) and (6) of the main text, the $T$-dependence of $1/\tau$ to leading order in the high-temperature expansion is
\begin{align}
\frac{1}{\tau}=\text{const}+\frac{aA}{T},
\end{align}
where 
$a=8\pi c mJ^4/3>0$ (the numerical coefficient $c$ depends on the lattice type)
and 
\begin{align}
A=\frac{1}{(2k_F)^3} \int^{2k_F}_0 dk\frac{k^2}{\sqrt{1-(k/2k_F)^2}}\left(\chi_k^0-\langle\chi\rangle\right).\label{A}
\end{align}
Here, $\langle\chi\rangle$ ia an average of $\chi^0_k$ over the Brillouin zone, which we model by a circle with radius $b$, 
chosen in such a way that its area coincides with that of the actual Brillouin zone. 
\begin{align}
\langle\chi\rangle=\frac{2}{b^2}\int^{b}_0 dk k \chi^0_k. 
\label{av}
\end{align}
 A particular choice of $b$ is irrelevant as long as $b\gg k_F$.
It is convenient to split the integral in Eq.~(\ref{av}) into two parts as $\int_0^{b} dk=\int^{2k_F}_{0}+\int_{2k_F}^{b}$. For $k_F\ll k\ll b$,
 $\chi^0_k$ falls off as  $\chi_1(k_F/k)^{2}$,  where
 $\chi_1\sim \chi^0_{2k_F}$. Therefore, the second integral can be estimated in the leading logarithmic approximation as 
\begin{align}
\frac{2}{b^2}\int^b_{2 k_F} dk k \chi_k^0 \approx 2\chi_1\left(\frac{k_F}{b}\right)^2\ln\frac{b}{k_F}. 
\end{align}
Combining the rest of $\langle\chi\rangle$ with the integral over $\chi^0_k$ in Eq.~(\ref{A}), we rewrite $A$ as
\begin{eqnarray}
A&=&\int^{2k_F}_0 \frac{dk k}{(2k_F)^2} \chi^0_k\left[\frac{k}{2k_F}\frac{1}{\sqrt{1-(k/2k_F)^2}}-2\pi\left(\frac{k_F}{b}\right)^2\right]\nonumber\\
&&-\frac{\pi}{2}\chi_1\left(\frac{k_F}{b}\right)^2\ln\frac{b}{k_F}.
\end{eqnarray}
Unless $\chi^0_k$ is peaked at some $k\ll k_F$, typical $k$ in the integral above is on the order of $k_F$. This means that the first term in the square brackets is on the order of one, while the second one 
is much smaller than one by the condition $k_F\ll b\sim 1$. The third term is small by the same condition. Therefore, $A$ can be approximated by the first term in the square bracket, which is positive definite, and the $T$ dependence of $1/\tau$ is of the insulating sign.

\section{Temperature dependence of $\rho_\text{RKKY}$}
The temperature dependence of $\rho_\text{RKKY}$ from Born approximation can be read out from Eq.~(6) and (8) in the main text:
\begin{align}
\rho_\text{RKKY} (T)&= \frac{4}{\pi} \rho_0 \int_0^1 dx \frac{x^2}{\sqrt{1-x^2}} \Big[ 1+K \tilde{\chi}_{2k_F x} \nonumber \\
&\quad + K^2 \left( \tilde{\chi}_{2k_F x}^2 - \langle \tilde{\chi}^2 \rangle  \right) \Big] \nonumber \\
& \approx \frac{a}{T-T^*}+b,\label{Eq:T0_defivation}
\end{align}
where
\begin{subequations}
\begin{align}
a &=  \frac{8J^2}{3\pi}\rho_0  \int_0^1 dx \frac{x^2}{\sqrt{1-x^2}} \tilde{\chi}_{2k_F x},  \\
b &=  \frac{4}{\pi} \rho_0 \int_0^1 dx \frac{x^2}{\sqrt{1-x^2}}, \\
T^*  &= \frac{2J^2}{3} \frac{ \int_0^1 dx \frac{x^2}{\sqrt{1-x^2}} \left( \tilde{\chi}_{2k_F x}^2 - \langle \tilde{\chi}^2 \rangle \right)}{ \int_0^1 dx \frac{x^2}{\sqrt{1-x^2}} \tilde{\chi}_{2k_F x}}.\label{Eq:T0_highT}
\end{align}
\end{subequations}

$T^*$ represents the effect of higher order corrections to the leading $1/T$ behavior that is obtained  from the high-T expansion. Thus, Eq.~\eqref{Eq:T0_defivation} is expected to hold as long as $T>>|T^*|$  . Consequently, if a fit based on Eq.~\eqref{Eq:T0_defivation} gives a value of $|T^*|$  higher than the position of the resistivity minimum, one should not rely on the high-$T$ expansion for computing the temperature dependence of the resistivity. A numerical approach, like the KPM-LD described in the main text, would still be appropriate.

\begin{figure}[!tbp]
	\includegraphics[width=\columnwidth]{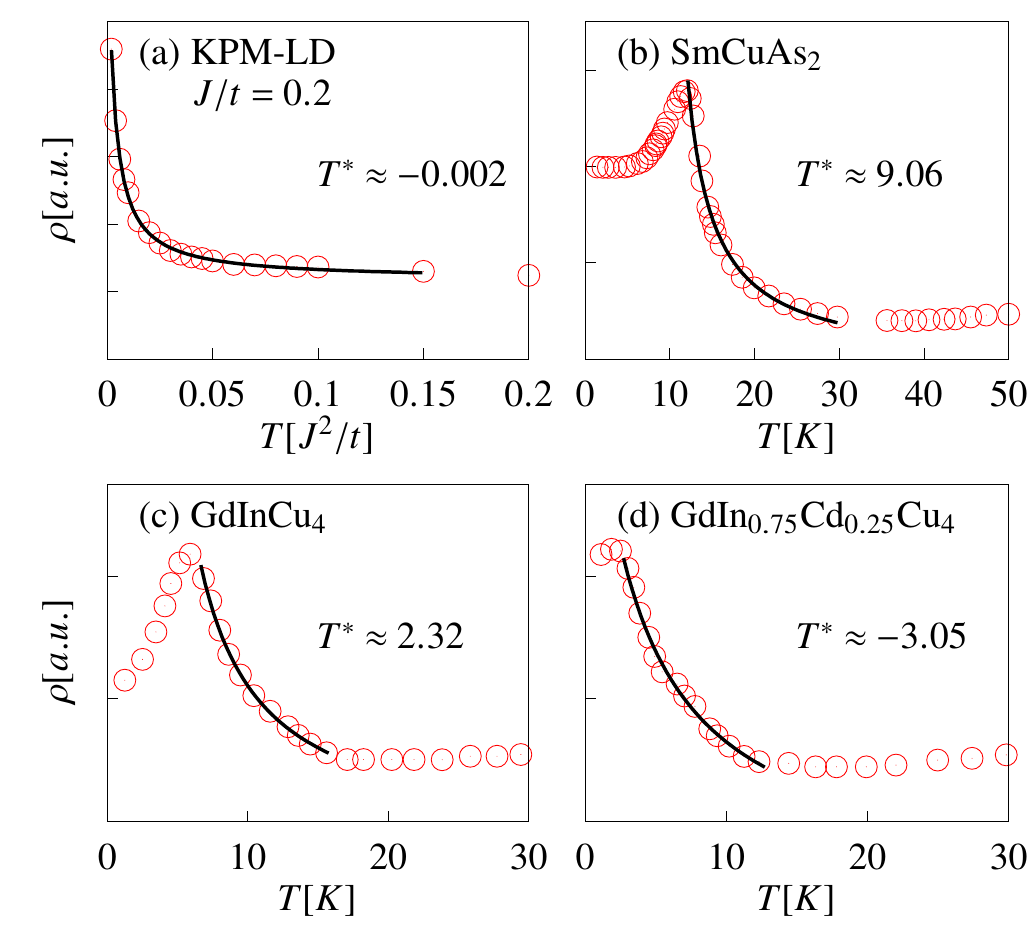}
	\caption{ \label{FigS1} Fits (solid lines) to various resistivity data (circles) by Eq.~\eqref{Eq:T0_defivation} with three parameters $\{a,b,T^* \}$. Data points are taken from: (a) KPM-LD results for triangular KLM with $J/t=0.2$ at filling $n=0.09$. (b) SmCuAs$_2$, from Ref.~\cite{Sampathkumaran2003}. (c-d) GdIn$_{1-x}$Cd$_x$Cu$_4$, from Ref.~\cite{Fritsch06}. } 
\end{figure}

Eq.~\eqref{Eq:T0_highT} yields $T^* \approx 0.01J^2/t$ for the triangular KLM with filling  fraction $n=0.09$. For general models with complicated Fermi surfaces, the sign of $T^*$ can be either positive or negative. 
For example, a fit to Eq.~\eqref{Eq:T0_defivation} of our KPM-LD data at $J/t=0.2$ gives $T^* \approx -0.002 J^2/t$
[see Fig.~\ref{FigS1}(a)]. To further test the validity of Eq.~\eqref{Eq:T0_defivation}, we also include in Figs.~\ref{FigS1}(b-d) the resistivity fits for a few specific materials, in which the Kondo effect is believed to be absent~\cite{Sampathkumaran2003, Fritsch06}. In all these cases, Fig.~\ref{FigS1} shows that the non-logarithmic resistivity upturns can be well described by Eq.~\eqref{Eq:T0_defivation}.  

\begin{figure}[!tbp]
	\includegraphics[width=0.9\columnwidth]{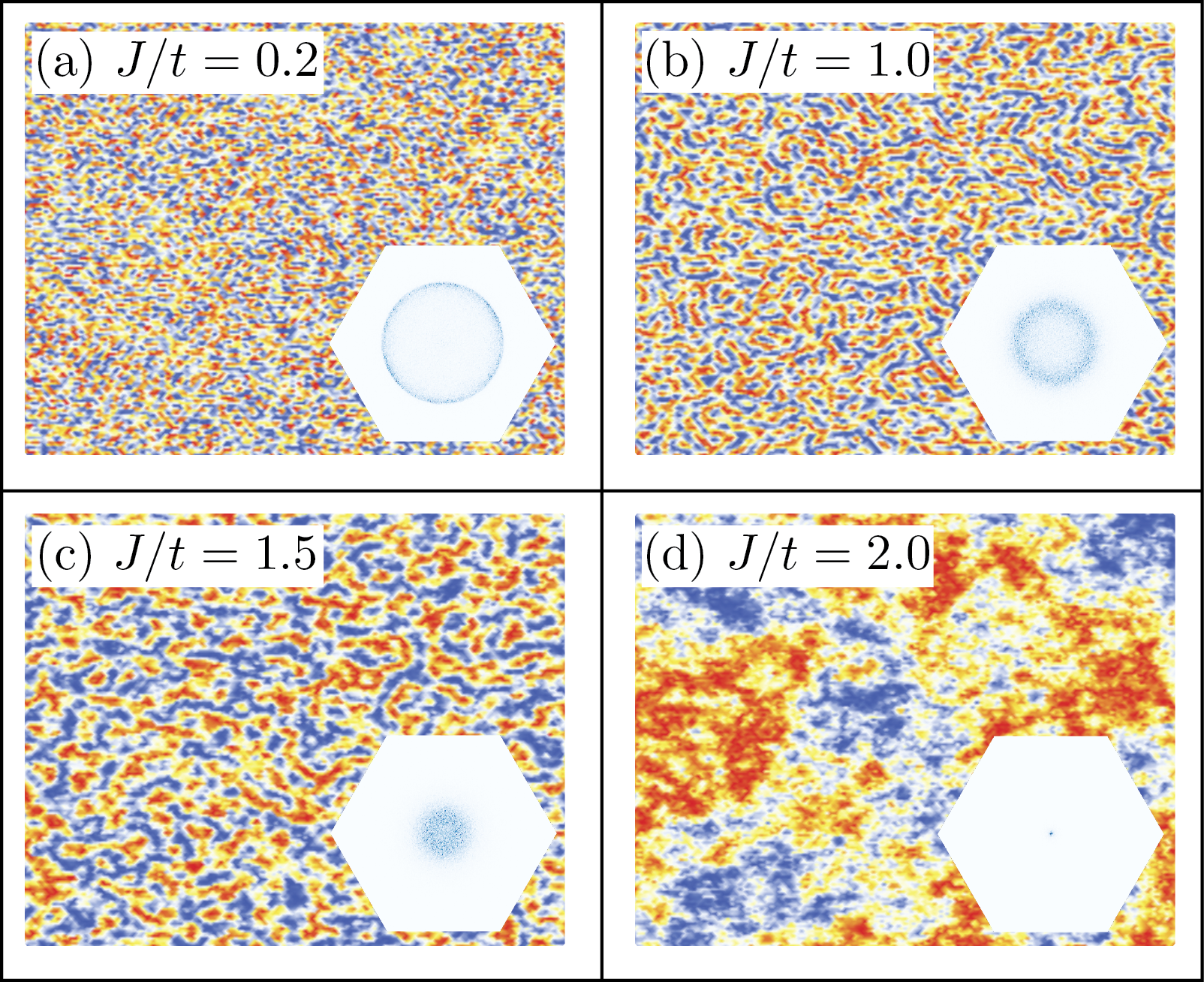}
	\includegraphics[width=0.9\columnwidth]{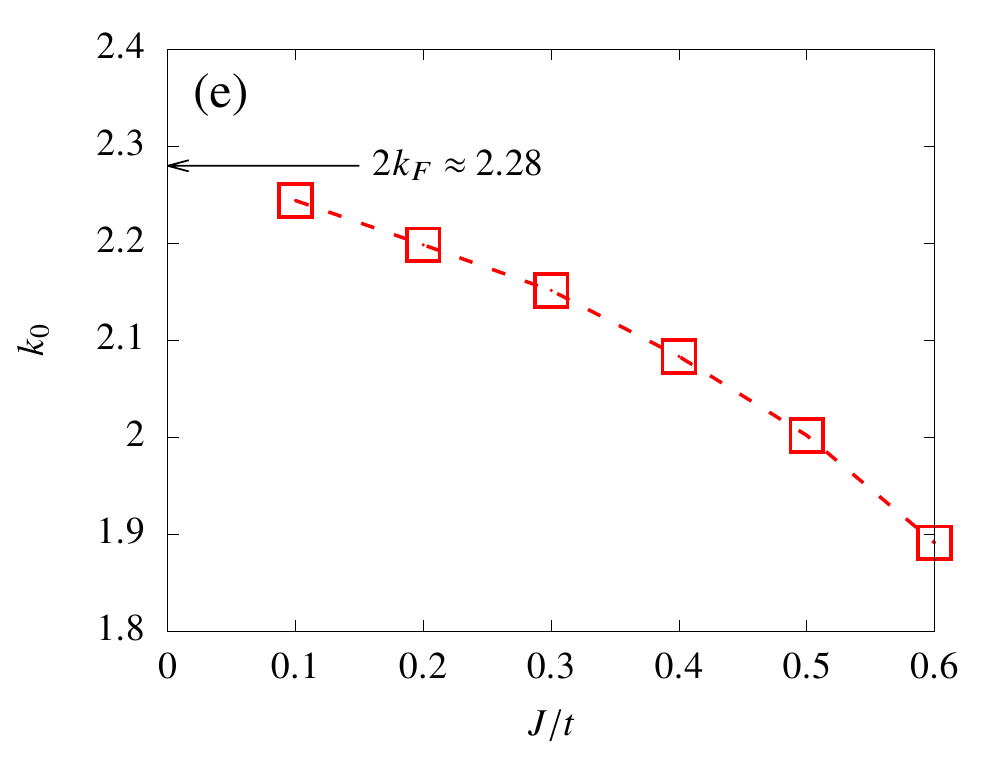}
	\caption{ \label{FigS2} (a-d) Real-space fluctuations obtained from KPM-LD for a triangular KLM at filling fraction $n=0.09$ and temperature $T=0.002 J^2/t$. The color gradient represents an arbitrary component of the spin 3-vector. Insets: corresponding $\mathcal{S}(\bm{k})$ in the 1st BZ. (e) KPM-LD results of momentum-space ring radius $k_0$ as a function of $J/t$, again with fixed parameters $n=0.09$ and temperature $T=0.002 J^2/t$. 
(the dashed line is guide to the eye)}
\end{figure}

\section{Spherical approximation}
In the spherical approximation, the local constraint, $|\bm{S}_i|=1$, is replaced  by a global one: $\sum_i |\bm{S}_i|^2=N$, where $N$ is the total number of sites. The partition function becomes:
\begin{align}
\mathcal{Z}&=\int \mathcal{D}\bm{S} d\lambda e^{-\beta \mathcal{H}-i \lambda \left( \sum_i |\bm{S}_i|^2-N \right)} \nonumber \\
&=\int \mathcal{D}\bm{S} d\lambda e^{-\beta \sum_{\bm{k}}\left[ \Delta - J^2\tilde{\chi}_{\bm{k}} \right] |S_{\bm{k}}|^2},
\end{align}
where $\lambda \equiv -i \beta \Delta$ is a Lagrangian multiplier introduced to enforce the global constraint.
After integrating out the spins, we obtain the structure factor:
\begin{equation}
{\cal S}( {\bm k} ) = \frac{3T}{2 [ \Delta(T) - J^2 {\tilde \chi}_{\bm k}]},
\end{equation}
where the parameter $\Delta(T)$ is determined by the self-consistency condition $\sum_{\bm{k}}|\bm{S}_{\bm{k}}|^2=\sum_i |\bm{S}_i|^2=N$, or equivalently
\begin{equation}
\frac{1}{N} \sum_{\bm k} \frac{J^2}{\Delta(T) - J^2 {\tilde \chi}_{\bm k}} = \frac{2 J^2 \beta}{3}.
\label{delta}
\end{equation}

\begin{figure}[!tbp]
	\includegraphics[width=\columnwidth]{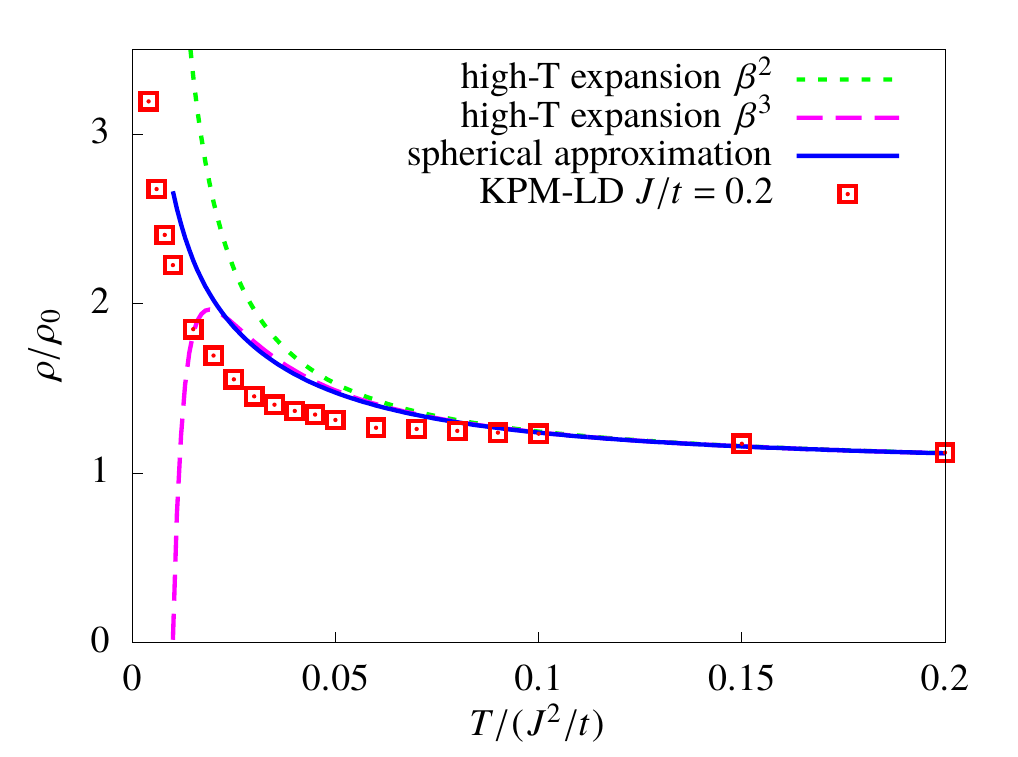}
	\caption{ \label{FigS3} Temperature dependence of the resistivity 
	for a triangular KLM at filling fraction $n=0.09$. 
	The lines correspond to calculations based on the the Born approximation and
	a bare susceptibility with a renormalized wave-vector, $\chi^0_k \to \chi^0_{ \alpha k} $, where the rescaling factor $\alpha$ 
	is chosen to  enforce the condition $k_0 \simeq 2.2$ for $J/t=0.2$ . The symbols correspond to the results of KPM-LD simulations.}
\end{figure}

\section{Renormalization of Fermi wave-vector}

In the weak-coupling limit $J/t \to 0$, the structure factor $\mathcal{S}(\bm{k})$ at low temperature is maximized at a ring with radius $k_0=2k_F$. However, as it is discussed in the main text, the ring shrinks towards the center of the 1st Brillouin zone upon increasing $J/t$.
Correspondingly, in real space, we see the liquid-like phase with characteristic wavelength $2\pi/k_0$, which finally develops into ferromagnetism at large couplings $J/t \gtrsim 2$ [see Figs.~\ref{FigS2} (a-d)].

To further illustrate this fact, we performed KPM-LD simulations on $256\times 256$ triangular lattice for $n=0.09$, $J/t=\{ 0.1,0.2,\ldots,0.6 \}$ and a relatively low temperature $T/(J^2/t)=0.002$. The Chebyshev polynomial expansion order is $M=1000$ ($J/t=0.1$) and $M=500$ ($J/t=\{ 0.2,0.3,\ldots,0.6 \}$). The number of random vectors is $R=512$ ($J/t=0.1$) and $R=128$ ($J/t=\{ 0.2,0.3,\ldots,0.6 \}$). The numbers of total Langevin steps are $\{1 \times 10^4,3\times 10^3\}$ ($J/t=\{ 0.1,0.2 \}$), and $5\times 10^3$ ($J/t=\{ 0.3,\ldots,0.6 \}$), each of duration $\Delta \tau =100$ ($J/t=\{ 0.1,0.2 \}$) and $\Delta \tau =10$ ($J/t=\{ 0.3,\ldots,0.6 \}$). Then starting from the $2 \times 10^3$th Langevin step, we average the value of $k_0$ for spin configurations differed by every $100$ Langevin steps.


As shown in Figs.~1(d)(f) of the main text,  $k_0=2k_F$ ($k_0 \approx 2.28$)  in the weak-coupling  limit, $J/t \to 0$. This result arises from the bare magnetic susceptibility maximum at $k=2k_F$  [see Fig.~1(b)]. 
However, Fig.~\ref{FigS2}(e) shows that $k_0$ decreases monotonically as a function of $J/t$. For $J/t=0.2$, the difference between $k_0$ and the bare $2k_F$ ($2k_F$ at $J/t=0$) is approximately equal to  $0.08$. This shift of $k_0$ as a
function of $J/t$ explains the low-temperature deviation between the analytical result for $\rho$, obtained in the weak-coupling
limit, and the numerical result obtained for $J/t=0.2$ [see Fig.~2(a)]. To better illustrate this point, 
Fig.~\ref{FigS3} shows a comparison between  the analytical results based on a bare susceptibility function in which the wave-vector
has been rescaled ($\chi^0_k \to \chi^0_{\alpha k} $) to enforce the condition $k_0 \simeq 2.2$ for $J/t=0.2$.
These clear improvement of the agreement between the analytical and the numerical result indicates that the shift 
in $k_0$, which is not captured at the RKKY level, is the main source of deviation between the resistivity curve, $\rho(T)$, for
finite $J/t$ and the curve obtained in the weak-coupling limit $J/t \to 0$.

\bibliographystyle{apsrev4-1}
\bibliography{ref}

\begin{thebibliography}{37}
\bibitem [{AA()}]{AA}%
  \BibitemOpen
  \href@noop {} {}\bibinfo {note} {The ``metallicity parameter''
  $\varepsilon_F\tau/\hbar$ is not particularly large in these compounds (for
  example, $\varepsilon_F\tau/\hbar\approx 13$ in
  Ref.~\onlinecite{Nakatsuji06}), and one could in principle invoke
  localization as an explanation of the resistivity upturn. However, the fact
  that the resistivity maximum occurs at the ordering temperature points to the
  magnetic nature of the effect.}\BibitemShut {Stop}%
\bibitem [{com()}]{comment_lowT}%
  \BibitemOpen
  \href@noop {} {}\bibinfo {note} {The data point at $J/t=2, T=0.002J^2/t$ is
  accurate to order $\sigma = 110 \pm 10 e^2/h$. The primary sources of error
  are the finite $M$ and the incomplete equilibration at this lowest
  temperature data point.}\BibitemShut {Stop}%
\end{thebibliography}

\end{document}